\documentclass[aps,twocolumn,showpacs,floatfix]{revtex4-1}
\usepackage{graphicx}
\usepackage{epsfig}
\usepackage{times}
\usepackage{amsmath,amssymb}
\usepackage{amsthm}
\usepackage{subfigure,overpic}
\usepackage{mathrsfs}
\usepackage{multirow}

\def \beq {\begin{equation}}
\def \edq {\end{equation}}
\def \bes {\begin{subequations}}
\def \eds {\end{subequations}}
\def \beqn {\begin{equation*}}
\def \edqn {\end{equation*}}

\def \dag {\dagger}

\def \up {\uparrow}
\def \down {\downarrow}

\def \calh {{\cal{H}}}

\def \ket {\rangle}
\def \bra {\langle}

\begin{document}
\title{Large thermoelectric power and figure of merit in a ferromagnetic-quantum dot-superconducting device}
\author{Sun-Yong Hwang}
\affiliation{Institut de F\'{\i}sica Interdisciplin\`aria i Sistemes Complexos
IFISC (CSIC-UIB), E-07122 Palma de Mallorca, Spain}
\author{Rosa L\'opez}
\affiliation{Institut de F\'{\i}sica Interdisciplin\`aria i Sistemes Complexos
IFISC (CSIC-UIB), E-07122 Palma de Mallorca, Spain}
\author{David S\' anchez}
\affiliation{Institut de F\'{\i}sica Interdisciplin\`aria i Sistemes Complexos
IFISC (CSIC-UIB), E-07122 Palma de Mallorca, Spain}

\begin{abstract}
We investigate the thermoelectric properties of a quantum dot coupled to ferromagnetic and superconducting electrodes. The combination of spin polarized tunneling at the ferromagnetic-quantum dot interface and the application
of an external magnetic field that Zeeman splits the dot energy level leads to large values of the thermopower
(Seebeck coefficient). Importantly, the thermopower can be tuned with an external gate voltage connected to the dot.
We compute the figure of merit that measures the efficiency of thermoelectric conversion and find that it attains
high values. We discuss the different contributions from Andreev reflection processes and quasiparticle tunneling into
and out of the superconducting contact.
Furthermore, we obtain dramatic variations of both the magnetothermopower and the spin Seebeck effect,
which suggest that in our device spin currents can be controlled with temperature gradients only.
\end{abstract}

\maketitle
\section{Introduction}

The quest for energy harvesting devices that efficiently convert waste heat into electricity has been intense
in the last decades. It has been argued that low dimensional systems offer better performances
due to interfacial boundary scattering of phonons~\cite{hic93} and strongly energy dependent transmissions~\cite{mah96}.
The generated thermopower is given by the Seebeck coefficient $S$ as the ratio between the created electric voltage
and the temperature difference applied across the device with vanishing net current~\cite{gol10}.
It turns out that electron-hole asymmetry present in the system determines the size of the thermoelectric effects given by $S$.
The reason is that the electron and hole thermocurrents generated in response to a thermal gradient flow in
opposite directions and for a particle-hole symmetric density of states (DOS) they exactly cancel each other,
thus giving zero thermovoltage.

On the other hand, even if the Seebeck coefficient is large a heat current inevitably accompanies a temperature
gradient. The efficiency of the thermoelectric process is then determined by the dimensionless figure of merit $ZT$, which
accounts for the relation between the thermoelectric power factor
and the heat or thermal conductance~\cite{iof57}. One would naively think that maximization of the figure of merit is expected in superconducting materials, which are perfect electric conductors and poor thermal conductors. However, the problem is that the superconducting DOS
exhibits electron-hole symmetry and hence the thermopower is strongly suppressed~\cite{gin91}.

Recent proposals suggest that electron-hole symmetry can be broken if the superconductor
is put in proximity of ferromagnetic contacts~\cite{mac13} or by combining an external magnetic field with a spin filter~\cite{oza14}.
The symmetry breaking originates from an exchange field induced splitting of the spin up and down energy subbands in the superconductor. 
As a consequence, large thermoelectric effects are predicted. The effect disappears if the spin polarization of the ferromagnetic
side of the junction vanishes. A very recent work reports the observation of
enhanced thermocurrents in superconductor-ferromagnet tunnel-coupled junctions~\cite{kol15}.
Similarly, large values of the Seebeck coefficient are found in superconducting-normal bilayers with spin active interfaces~\cite{kal14}
or if layered structures are considered~\cite{mac14}.

\begin{figure}[t]
\centering
  \begin{centering}
    \includegraphics[width=0.138\textwidth,clip,angle=90]{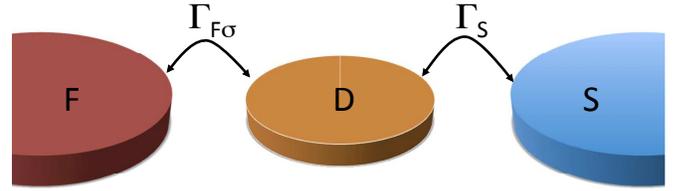}
  \end{centering}
  \caption{(Color online) Schematic illustration of our device. A quantum dot (D) is sandwiched
  between the left electrode, which is hot and ferromagnetic (F), and the right reservoir, which is cold and
  superconducting (S).   The tunnel couplings are indicated. We note that the hybridization between
  the source contact and the dot is spin dependent with $\sigma=\{\uparrow,\downarrow\}$. The energy level of the quantum dot
  can be tuned with a capacitively coupled gate terminal (not shown here).}
  \label{fig:sketch}
\end{figure}

Here, we propose to insert a quantum dot between the ferromagnetic source electrode and the superconducting drain contact
as in Fig.~\ref{fig:sketch}.
The advantage of the setup lies on the easy manipulation of the electron-hole symmetry in the local DOS by electrically coupling
the dot to a nearby gate. The thermopower thus acquires a characteristic sawtooth structure as the gate potential sweeps across
resonances in a semiconductor dot attached to normal leads~\cite{sta93,fah13}.
The gate voltage also controls the electron number in the dot. Furthermore, the sharp
resonances in the dot allow us to play with energy filtering effects
that may lead to an optimal thermoelectric conversion~\cite{hum05}.
We find that the thermopower can be tuned between $0$ and $350$~$\mu$V/K
when the energy level of the quantum dot is varied around the Fermi energy in the scale
of the superconducting gap.
Consequently, the figure of merit $ZT$ increases from $0$ up to $6$.

The hybrid system considered here is also interesting for fundamental reasons.
It is well known that at low bias voltages the dominant transport mechanism at the interface
between a metal and a superconductor is an electron-hole conversion known as Andreev reflection~\cite{tin96}.
The Andreev reflection generates a Cooper pair in the superconducing condensate and as such involves a spin flip process.
Therefore, when the metal is ferromagnetic
one expects a strong impact in the current--voltage characteristic of a ferromagnetic-superconducting junction
as a function of the magnetization distribution~\cite{jon95,aum00}.
One of the most spectacular effects is the appearance of drag currents in three-terminal structures
as a result of cross Andreev processes~\cite{deu00,fal01,mel04,bec04}.
Insertion of a large quantum dot increases the strength of the effect~\cite{san03}.
Additionally, ferromagnetic-quantum dot-superconducting systems serve as excellent
platforms to study spin correlation effects~\cite{hof10}.

Yet, we~\cite{hwa15} recently pointed out that the Andreev current of a normal-quantum dot-superconducting device is zero to all orders in the temperature shift and therefore Andreev processes cannot alone generate finite thermovoltages due to the Onsager reciprocity relation~\cite{cla96,jac12}.
This also holds for a ferromagnetic contact, as we show below. Hence, it is crucial to take into account
the role of quasiparticle currents. These appear when single electrons fill states in the source contact
and can tunnel to empty quasiparticle states in the drain. The quasiparticle current [see Eq.~\eqref{I_Q} below] depends on the quasiparticle transmission [see Eq.~\eqref{eq:TQ} below], which is a function of superconducting density of states and changes abruptly on the scale of $\Delta$. Therefore, we need Zeeman field splittings of the order of $\Delta$ to create abrupt changes in the thermopower. In addition, we observe enhanced values of the Seebeck coefficient
for ferromagnet polarizations close to the half-metallic case because the Andreev conductance term
sharply decreases when the polarization approaches $100$~\% and the thermopower thus increases~\cite{mac13}.
The sensitivity of $S$ to the polarization is also reflected in the magnetothermopower, which shows
strong variations as a function of the energy level and the Zeeman splitting upon reversal
of the magnetization. Remarkably,
we find very large values of $ZT$ as a result of the combined effect of external magnetic fields, spin polarization
and appropriate tuning of the dot energy level.

We also explore spin caloritronic effects that arise when a spin polarized current is generated under the influence
of a temperature bias~\cite{uch08,sla10,jaw10}. For Andreev processes we below demonstrate that the spin current vanishes
even if Zeeman splittings or spin-dependent tunnel couplings are present. This is valid even in nonequilibrium conditions
(finite voltage and temperature biases) and is due to a symmetry between the electron and hole sectors. However, the quasiparticle
current is free from this constraint and we expect, quite generally, nonzero spin currents in temperature driven junctions.
A way to quantify thermally generated spin voltages is to define the spin Seebeck coefficient $S_s$. It is natural that
when the ferromagnet polarization and the external magnetic field are zero the spin thermopower vanishes.
Nevertheless, we obtain values of $S_s$ as large as $500$~$\mu$V/K in the presence of Zeeman splittings
and nonzero magnetization. Our findings thus suggest that a ferromagnetic-quantum dot-superconducting device
is a good candidate to create large spin population imbalances using thermal means only.


The paper is organized as follows. In Sec.~\ref{theory}, we explain the theoretical model. The magnetization in the ferromagnetic contact leads to spin-dependent tunneling rates for the ferromagnetic-quantum dot coupling.
We discuss both the electric current and the heat flux driven by voltage or temperature biases and separate the contributions
from Andreev processes and quasiparticle tunneling. In Sec.~\ref{linear} we give the transport coefficients (electric, thermoelectric, electrothermal and thermal conductances) that characterize the linear transport regime. We show that the cross responses obey the Kelvin-Onsager relation. We also derive appropriate expressions for the charge thermopower, the figure of merit and the spin Seebeck coefficient. Section~\ref{results} contains our main results.
We discuss the dependence of the thermoelectric
effects for both charge and spin on the spin polarization, the applied magnetic field and the external gate voltage. 
Finally, we summarize our findings in Sec.~\ref{conclusion}.

\section{Hamiltonian and Green's Function Approach}\label{theory}
The F-D-S device is comprised of the left ferromagnet (F) with a polarization $p$ ($|p|\le1$), a single-level quantum dot (D), and the right superconductor (S) as depicted in Fig. 1. The total Hamiltonian reads
\begin{equation}
\calh=\calh_{F}+\calh_{S}+\calh_{D}+\calh_{T}\,,
\end{equation}
where
\begin{equation}
\calh_{F}=\sum_{k\sigma}\varepsilon_{Fk\sigma}c_{Fk\sigma}^{\dag}c_{Fk\sigma}
\end{equation}
describes the charge carrier with momentum $k$, spin $\sigma$ with a magnetization $M\sigma$ along the given axis (say $z$) in the left ferromagnet and
\begin{equation}
\calh_{S}=\sum_{k\sigma}\varepsilon_{Sk\sigma}c_{Sk\sigma}^{\dag}c_{Sk\sigma}+\sum_{k}[\Delta c_{S,-k\up}^{\dag}c_{Sk\down}^{\dag}+{\rm H.c.}]
\end{equation}
describes the right superconductor with an order parameter given by the energy gap $\Delta$. We neglect the phase of $\Delta$ and treat it as a real constant. This is valid with a suitable gauge transformation when we consider an equilibrium superconductor. In what follows, we set $\Delta=1$ as an energy unit. In the dot Hamiltonian
\begin{equation}
\calh_{D}=\sum_{\sigma}\varepsilon_{d\sigma}d_{\sigma}^{\dag}d_{\sigma}\,,
\end{equation}
the spin-degenerate energy level can be split when the magnetic field is on, i.e., $\varepsilon_{d\sigma}=\varepsilon_d+\sigma\Delta_Z$ with Zeeman splitting $\Delta_Z$. Scattering at the ferromagnet creates spin-dependent scattering phases that to first order induce an additional effective Zeeman splitting and can be included into $\Delta_Z$~\cite{cot06}.
Finally, the charge tunneling between the quantum dot and each lead is described by
\begin{equation}
\calh_{T}=\sum_{k\sigma}t_{F\sigma}c_{Fk\sigma}^{\dag}d_{\sigma}+\sum_{k\sigma}t_{S\sigma}c_{Sk\sigma}^{\dag}d_{\sigma}+{\rm H.c.}
\end{equation}

We ignore spin-flip scattering  (see, however, Ref.~\cite{cao04}) and evaluate the spin-resolved charge and heat currents from the time evolution of spin-$\sigma$ electron number $N_{F\sigma}=\sum_{k}c_{Fk\sigma}^{\dag}c_{Fk\sigma}$ and the energy $\calh_{F\sigma}=\sum_{k}\varepsilon_{Fk\sigma}c_{Fk\sigma}^{\dag}c_{Fk\sigma}$ in the left ferromagnet
\begin{align}
&I_\sigma=-(ie/\hbar)\bra[\calh,N_{F\sigma}]\ket\,,\\
&J_\sigma=-(i/\hbar)\bra[\calh,\calh_{F\sigma}]\ket-I_\sigma V\,,
\end{align}
where the last term corresponds to the Joule heating.
Applying the nonequilibrium Keldysh-Green formalism~\cite{Cue96,Sun99}, we find that the current for each spin $I_\sigma=I_{A}^{\sigma}+I_{Q}^{\sigma}$ is given by a sum of two terms, i.e., the spin-resolved Andreev current $I_{A}^{\sigma}$ and that of quasiparticle contribution $I_{Q}^{\sigma}$ in terms of their corresponding transmission probabilities $T_A^\sigma$ and $T_Q^\sigma$,
\begin{align}
&I_{A}^{\sigma}=\frac{e}{h}\int d\varepsilon~T_{A}^{\sigma}(\varepsilon)\big[f_{F}(\varepsilon-eV)-f_{F}(\varepsilon+eV)\big]\label{I_A}\,,\\
&I_{Q}^{\sigma}=\frac{e}{h}\int d\varepsilon~T_{Q}^{\sigma}(\varepsilon)\big[f_{F}(\varepsilon-eV)-f_{S}(\varepsilon)\big]\label{I_Q}\,,
\end{align}
where $f_{\alpha=F,S}(\varepsilon\pm eV)=\{1+\exp[(\varepsilon\pm eV-E_F)/k_{B}T_{\alpha}]\}^{-1}$ is the Fermi-Dirac distribution with the applied voltage $V=V_F-V_S$ and the electrode temperature $T_\alpha=T+\theta_\alpha$ ($T$: background temperature, $\theta_\alpha$: thermal bias). For definiteness, we drive only the F lead assuming that the right superconductor is at equilibrium ($V_S=\theta_S=0$) and take the Fermi level as the energy reference ($E_F=0$).
Similarly, the heat current is given by $J_\sigma=J_{A}^{\sigma}+J_{Q}^{\sigma}$ with
\begin{align}
&J_{A}^{\sigma}=-2VI_{A}^{\sigma}\label{J_A}\,,\\
&J_{Q}^{\sigma}=\frac{1}{h}\int d\varepsilon~(\varepsilon-eV)~T_{Q}^{\sigma}(\varepsilon)\big[f_{F}(\varepsilon-eV)-f_{S}(\varepsilon)\big]\label{J_Q}\,.
\end{align}
In Eq.~\eqref{J_A}, the Andreev energy flow cancels out due to the particle-hole ($p$-$h$) symmetry and only the Joule part survives.
In other words, we have
\begin{equation}
J_{A\sigma}^{(p)}=\frac{1}{h}\int d\varepsilon~(\varepsilon-eV)~T_{A}^{\sigma}(\varepsilon)[f_{F}(\varepsilon-eV)-f_{F}(\varepsilon+eV)]\,,
\end{equation}
and
\begin{equation}
J_{A\sigma}^{(h)}=\frac{1}{h}\int d\varepsilon~(\varepsilon+eV)~T_{A}^{\sigma}(\varepsilon)[f_{F}(\varepsilon+eV)-f_{F}(\varepsilon-eV)]\,.
\end{equation}
Hence $J_{A}^{\sigma}=J_{A\sigma}^{(p)}+J_{A\sigma}^{(h)}=-2VI_{A}^{\sigma}$. Importantly, this property causes the Andreev heat current  to vanish in the linear response regime when we apply a small voltage bias (i.e., no Peltier effect). Note that the factor 2 in Eq.~\eqref{J_A} comes from an equal contribution of particle and hole to the heat current.

The key quantities to determine the transmission probabilities $T_{A}^{\sigma}$ and $T_{Q}^{\sigma}$ are the dot retarded Green's functions $G^r_{ij}(\varepsilon)$ ($i,j=1,2,3,4$) calculated in the spin-generalized Nambu spinor basis $\widehat{d}=(d_\up,d_\down^\dag,d_\down,d_\up^\dag)^T$
\beq\label{Green}
{\bf G}^r_{d}(\varepsilon)
=\left(\begin{array}{cc|cc}
G^{r}_{11}(\varepsilon)& G^{r}_{12}(\varepsilon)& 0& 0\\
G^{r}_{21}(\varepsilon)& G^{r}_{22}(\varepsilon)& 0& 0\\
\hline
0& 0& G^{r}_{33}(\varepsilon) &G^{r}_{34}(\varepsilon)\\
0& 0&G^{r}_{43}(\varepsilon) & G^{r}_{44}(\varepsilon)\\
\end{array}\right)\,,
\edq
where $2\times2$ submatrices in the first block ($i,j=1,2$) and the second block ($i,j=3,4$) correspond to the electron spin-up and spin-down spaces respectively, with subscripts 1, 3 denoting electron sectors and 2, 4 referring to hole parts. The whole matrix in Eq.~\eqref{Green} is block-diagonal since we have ignored spin-flip processes, which thus separates the spin spaces. The Green's functions are explicitly given by~\cite{cao04,Sun99}
\begin{align}
&G^{r}_{11}=\Big[\varepsilon-\varepsilon_{d\up}+\frac{i\Gamma_{F\up}}{2}+\frac{i\Gamma_{S}}{2}\beta_d(\varepsilon)+\frac{\Gamma_S^{2}\Delta^{2}A_1^r(\varepsilon)}{4(\varepsilon^2-\Delta^{2})}\Big]^{-1},\label{G11}\\
&G^{r}_{33}=\Big[\varepsilon-\varepsilon_{d\down}+\frac{i\Gamma_{F\down}}{2}+\frac{i\Gamma_{S}}{2}\beta_d(\varepsilon)+\frac{\Gamma_S^{2}\Delta^{2}A_2^r(\varepsilon)}{4(\varepsilon^2-\Delta^{2})}\Big]^{-1},\label{G33}\\
&G^{r}_{12}=G^{r}_{21}=G^{r}_{11}\frac{i\Gamma_{S}}{2}\beta_o(\varepsilon)A_1^r(\varepsilon)\,,\\
&G^{r}_{34}=G^{r}_{43}=-G^{r}_{33}\frac{i\Gamma_{S}}{2}\beta_o(\varepsilon)A_2^r(\varepsilon)\,,
\end{align}
with
\begin{align}
&A_1^r(\varepsilon)=\Big[\varepsilon+\varepsilon_{d\down}+\frac{i\Gamma_{F\down}}{2}+\frac{i\Gamma_{S}}{2}\beta_d(\varepsilon)\Big]^{-1},\\
&A_2^r(\varepsilon)=\Big[\varepsilon+\varepsilon_{d\up}+\frac{i\Gamma_{F\up}}{2}+\frac{i\Gamma_{S}}{2}\beta_d(\varepsilon)\Big]^{-1},\\
&\beta_d(\varepsilon)=\frac{\Theta(|\varepsilon|-\Delta)|\varepsilon|}{\sqrt{\varepsilon^{2}-\Delta^{2}}}-i\frac{\Theta(\Delta-|\varepsilon|)\varepsilon}{\sqrt{\Delta^{2}-\varepsilon^{2}}}\,,\\
&\beta_o(\varepsilon)=\frac{\Theta(|\varepsilon|-\Delta)\text{sgn}(\varepsilon)\Delta}{\sqrt{\varepsilon^{2}-\Delta^{2}}}-i\frac{\Theta(\Delta-|\varepsilon|)\Delta}{\sqrt{\Delta^{2}-\varepsilon^{2}}}\,.
\end{align}
The remaining Green's functions follow from Eqs.~\eqref{G11} and \eqref{G33}: $G_{22}^{r}(\varepsilon)=-G_{33}^{r,*}(-\varepsilon)$ and $G_{44}^{r}(\varepsilon)=-G_{11}^{r,*}(-\varepsilon)$. We have used the wide band approximation, i.e., energy-independent tunnel couplings, $\Gamma_{F\sigma}=\Gamma_F(1+\sigma p)=2\pi|t_{F\sigma}|^2 \sum_k \delta (\varepsilon-\varepsilon_{Fk\sigma})$ and $\Gamma_{S\sigma}=\Gamma_S=2\pi|t_{S\sigma}|^2\sum_{p}\delta(\varepsilon-\varepsilon_{Sp\sigma})$ with $\Gamma_{\alpha}=(\Gamma_{\alpha\up}+\Gamma_{\alpha\down})/2$ being the spin-averaged coupling constant to each lead $\alpha=F,S$.
The spin dependence in  $\Gamma_{F\sigma}$ arises from the nonzero magnetization in the ferromagnet,
\begin{equation}
p=\frac{\nu_\uparrow-\nu_\downarrow}{\nu_\uparrow+\nu_\downarrow}\,,
\end{equation}
 where $\nu_\sigma=\sum_k \delta (\varepsilon-\varepsilon_{Fk\sigma})$
is the ferromagnet density of states.
We wish to point out that for $p=\Delta_Z=0$ the spin dependence of the system disappears
as we would end up with a normal-quantum dot-superconducting setup~\cite{gab04,dea10}.
Possible spintronic (or spin caloritronic) nature can only emerge with a nonzero polarization or a magnetic field, or the combination of both, either $p\ne0$ or $\Delta_Z\ne0$.

One can write the spin-resolved Andreev transmission in terms of Green's functions, viz. 
\begin{subequations}\label{eq:TA}
\begin{align}
&T_{A}^{\up}(\varepsilon)=\Gamma_{F\up}\Gamma_{F\down}\big|G_{12}^{r}(\varepsilon)\big|^{2}\,,\\
&T_{A}^{\down}(\varepsilon)=\Gamma_{F\down}\Gamma_{F\up}\big|G_{34}^{r}(\varepsilon)\big|^{2}\,,
\end{align}
\end{subequations}
with which the Andreev charge $I_A^{c}=I_{A}^{\up}+I_{A}^{\down}$ and spin $I_A^{s}=I_{A}^{\up}-I_{A}^{\down}$ currents can be defined via Eq.~\eqref{I_A}. The Andreev heat flux $J_A^c=-2VI_A^{c}$ and the spin-polarized one $J_A^s=-2VI_A^{s}$ can be determined from Eq.~\eqref{J_A}. 
Analogously, the quasiparticle charge and spin currents (and those of heat) are respectively given by $I_Q^{c}=I_{Q}^{\up}+I_{Q}^{\down}$ ($J_Q^{c}=J_{Q}^{\up}+J_{Q}^{\down}$) and $I_Q^{s}=I_{Q}^{\up}-I_{Q}^{\down}$ ($J_Q^{s}=J_{Q}^{\up}-J_{Q}^{\down}$) with the aid of Eqs.~\eqref{I_Q} and \eqref{J_Q} and the corresponding transmissions
\begin{subequations}\label{eq:TQ}
\begin{align}
&T_{Q}^{\up}(\varepsilon)=\Gamma_{F\up}\widetilde{\Gamma}_S\Big(\big|G_{11}^{r}\big|^{2}+\big|G_{12}^{r}\big|^{2}-\frac{2\Delta}{|\varepsilon|}\text{Re}\big[G_{11}^rG_{12}^{r,*}\big]\Big)\,,\\
&T_{Q}^{\down}(\varepsilon)=\Gamma_{F\down}\widetilde{\Gamma}_S\Big(\big|G_{33}^{r}\big|^{2}+\big|G_{34}^{r}\big|^{2}+\frac{2\Delta}{|\varepsilon|}\text{Re}\big[G_{33}^rG_{34}^{r,*}\big]\Big)\,,
\end{align}
\end{subequations}
where $\widetilde{\Gamma}_S=\Gamma_{S}\Theta(|\varepsilon|-\Delta)|\varepsilon|/\sqrt{\varepsilon^{2}-\Delta^{2}}$.

For $p=\Delta_Z=0$, one can easily show that $G_{11}^{r}(\varepsilon)=G_{33}^{r}(\varepsilon)$ and $G_{12}^{r}(\varepsilon)=-G_{34}^{r}(\varepsilon)$, hence $T_{A}^{\up}(\varepsilon)=T_{A}^{\down}(\varepsilon)$ and $T_{Q}^{\up}(\varepsilon)=T_{Q}^{\down}(\varepsilon)$ from Eqs.~\eqref{eq:TA} and \eqref{eq:TQ} respectively. In this case all the spin currents vanish identically, $I_A^s=I_Q^s=J_A^s=J_Q^s=0$, as expected. If either $p\ne0$ or $\Delta_Z\ne0$, the spin symmetry is generally broken, i.e., $T_{A}^{\up}(\varepsilon)\ne T_{A}^{\down}(\varepsilon)$ and $T_{Q}^{\up}(\varepsilon)\ne T_{Q}^{\down}(\varepsilon)$, leading to a spin-polarized net current. Nevertheless, focusing on Andreev transport only, the inherent particle-hole symmetry strictly satisfies $T_{A}^{\up}(\varepsilon)=T_{A}^{\down}(-\varepsilon)$ even in nonequilibrium conditions. Then, it follows from Eqs.~\eqref{I_A} and \eqref{J_A} that $I_A^s=J_A^s=0$ due to the symmetry of integrands in energy space, i.e.,
\begin{equation}
I_A^s=\frac{e}{h}\int d\varepsilon [T_{A}^{\up}(\varepsilon)-T_{A}^{\down}(\varepsilon)][f_{F}(\varepsilon-eV)-f_{F}(\varepsilon+eV)]=0\,,
\end{equation}
 with the property $f_{F}(-\varepsilon\pm eV)=1-f_F(\varepsilon\mp eV)$ (recall that we take $E_F=0$).
Therefore, in our model the subgap Andreev process always prohibits the generation of spin-polarized currents. 
Spin dependence of the crossed Andreev transport can be observed in the strong coupling regime with a multiterminal device~\cite{tro15} but here we consider a two-terminal device.

In stark contrast, $I_Q^s$ and $J_Q^s$ are generally nonzero with a finite $p$ or $\Delta_Z$ [see Eqs.~\eqref{I_Q}, \eqref{J_Q}, and \eqref{eq:TQ}]. Thus, net spin currents arise only from the quasiparticle contributions $I_s=I_Q^s$ and $J_s=J_Q^s$.
On the other hand, the total charge current $I_c=I_\up+I_\down$ and the total heat flux $J_c=J_\up+J_\down$ in general consist of the sum of both the Andreev parts and the quasiparticle contributions. Hence, we can write without loss of generality
\begin{align}
&I_c=I_A^c+I_Q^c\,,\\
&I_s=I_Q^s\,,\label{I_S}
\end{align}
and for the heat
\begin{align}
&J_c=J_A^c+J_Q^c\,,\label{J_C}\\
&J_s=J_Q^s\,.\label{J_S}
\end{align}

\section{Linear thermoelectric transport}\label{linear}
As can be seen in Eq.~\eqref{I_A}, the spin-resolved Andreev current $I_A^\sigma$ has no response to a thermal driving in the isoelectric case $V=0$~\cite{hwa15}, and in the linear regime we only have the contribution $I_A^\sigma=G_A^\sigma V$ with zero thermoelectric conductance. Accordingly, the Andreev heat flux $J_A^\sigma$ in Eq.~\eqref{J_A} has no linear term (the first nonzero term is quadratic). Thus, in linear response the spin-resolved charge and heat currents become
\begin{align}
&I_\sigma=(G_{A}^{\sigma}+G_{Q}^{\sigma})V+L_{Q}^{\sigma}\theta\,,\label{I_sigma}\\
&J_\sigma=R_{Q}^{\sigma}V+K_{Q}^{\sigma}\theta\,,\label{J_sigma}
\end{align}
with the corresponding transport coefficients given by
\begin{align}
&G_{A}^{\sigma}=\frac{2e^{2}}{h}\int d\varepsilon \big(-\partial_\varepsilon f\big)T_{A}^{\sigma}\,,\\
&G_{Q}^{\sigma}=\frac{e^{2}}{h}\int d\varepsilon\big(-\partial_\varepsilon f\big) T_{Q}^{\sigma}\,,\\
&L_{Q}^{\sigma}=\frac{e}{h}\int d\varepsilon\frac{\varepsilon-E_{F}}{T}\big(-\partial_\varepsilon f\big)T_{Q}^{\sigma}\,,\label{LQ}\\
&R_{Q}^{\sigma}=\frac{e}{h}\int d\varepsilon (\varepsilon-E_F)\big(-\partial_\varepsilon f\big) T_{Q}^{\sigma}\,,\label{RQ}\\
&K_{Q}^{\sigma}=\frac{1}{h}\int d\varepsilon \frac{(\varepsilon-E_F)^2}{T}\big(-\partial_\varepsilon f\big) T_{Q}^{\sigma}\,,
\end{align}
where $\partial_\varepsilon f$ denotes the energy derivative of Fermi function at equilibrium ($V_\alpha=\theta_\alpha=0$).
Equation~\eqref{J_C} further reduces to $J_c=J_Q^c$ as there is no Andreev contribution in Eq.~\eqref{J_sigma} which only appears in the nonlinear transport regime [Eq.~\eqref{J_A}]. Moreover, this vanishing linear Andreev heat current with $V$ is fundamentally linked to the Kelvin-Onsager relation, implying that the absence of $L_A^\sigma$ in Eq.~\eqref{I_sigma} due to the inherent particle-hole symmetry also guarantees the absence of $R_A^\sigma$ in Eq.~\eqref{J_sigma}.
For the quasiparticle coefficients, we find $R_Q^\sigma(p,\Delta_Z)=T L_Q^{\bar{\sigma}}(-p,-\Delta_Z)$
since the transmission obeys the relation $T_{A,Q}^\sigma(p,\Delta_Z)=T_{A,Q}^{\bar{\sigma}}(-p,-\Delta_Z)$ due to microreversibility.
This also implies $X^\sigma(p,\Delta_Z)=X^{\bar{\sigma}}(-p,-\Delta_Z)$ for all the kinetic coefficients $X=G,L,R,K$. Furthermore,
we obtain $R_Q^\sigma=T L_Q^\sigma$ as one can easily verify from Eqs.~\eqref{LQ} and~\eqref{RQ}. 

Employing Eqs.~\eqref{I_sigma} and \eqref{J_sigma}, we write the linear response charge and heat currents as
\begin{align}
&I_{c}=\sum_\sigma\Big[(G_{A}^{\sigma}+G_{Q}^{\sigma})V+L_{Q}^{\sigma}\theta\Big]\,,\label{IC}\\
&J_{c}=\sum_\sigma\Big[R_{Q}^{\sigma}V+K_{Q}^{\sigma}\theta\Big]\,,\label{JC}
\end{align}
and the spin-polarized counterparts as
\begin{align}
&I_{s}=(G_{Q}^{\up}-G_{Q}^{\down})V+(L_{Q}^{\up}-L_{Q}^{\down})\theta\,,\label{IS}\\
&J_{s}=(R_{Q}^{\up}-R_{Q}^{\down})V+(K_{Q}^{\up}-K_{Q}^{\down})\theta\,,\label{JS}
\end{align}
where we have used the symmetry relation $G_A^\up=G_A^\down$ in Eq.~\eqref{IS}. Notice that Eqs.~\eqref{IS} and \eqref{JS} are consistent with Eqs.~\eqref{I_S} and \eqref{J_S} when the linear response regime is considered. 

The linear responses found above completely determine the thermoelectric properties of our device.
Let us first focus on the charge transport.
The Seebeck coefficient or thermopower is defined as the generated voltage
from thermal gradients in open circuit conditions $I_c=0$.
This can be easily evaluated from Eq.~\eqref{IC}:
\beq\label{S}
S=-\frac{V}{\theta}\bigg|_{I_c=0}=\frac{\sum_\sigma L_{Q}^\sigma}{\sum_\sigma(G_{A}^{\sigma}+G_{Q}^{\sigma})}\,.
\edq

The efficiency of the thermoelectric conversion can be quantified by the figure of merit $ZT$.
We first calculate the thermal conductance with the help of Eqs.~\eqref{IC} and \eqref{JC}:
\beq\label{kappa}
\kappa=\frac{J_c}{\theta}\bigg|_{I_c=0}=\sum_\sigma K_{Q}^\sigma-\frac{1}{T}\frac{(\sum_\sigma R_{Q}^\sigma)^2}{\sum_\sigma(G_{A}^{\sigma}+G_{Q}^{\sigma})}\,,
\edq
Then, we find
\beq\label{ZT}
ZT=\frac{GS^2T}{\kappa}=\frac{\sum_\sigma(G_{A}^{\sigma}+G_{Q}^{\sigma})S^2T}{\sum_\sigma K_{Q}^\sigma-TS\sum_\sigma L_{Q}^\sigma}\,,
\edq
where we have used Eq.~\eqref{S} and the Kelvin-Onsager relation in order to rewrite Eq.~\eqref{kappa}. 
This expression clearly shows that a way to enhance the value of $ZT$ is to get higher $S$.
In deriving Eq.~\eqref{kappa} we have assumed that energy is carried by electronic degrees of freedom
only. We thus disregard the role of phonons, which can be nonnegligible at intermediate values
of the background temperature $T$.

We now turn to the spin-dependent transport. Quite generally, spin-dependent tunneling due to $\Gamma_{F\sigma}$
leads to spin accumulations in the F side. It will spin split the chemical potential of the magnetic
reservoir. If the size of the electrode is not sufficiently large and the spin-relaxation time
is long, we will have a nonzero spin bias $V_s$. Then, Eqs.~\eqref{IC} and \eqref{IS} are generalized as (let $G_{\sigma}=G_{A}^{\sigma}+G_{Q}^{\sigma}$)
\begin{align}
&I_{c}=(G_{\up}+G_{\down})V+\frac{1}{2}(G_{\up}-G_{\down})V_s+(L_{Q}^{\up}+L_{Q}^{\down})\theta\,,\\
&I_{s}=(G_{\up}-G_{\down})V+\frac{1}{2}(G_{\up}+G_{\down})V_s+(L_{Q}^{\up}-L_{Q}^{\down})\theta\,.
\end{align}
From these expressions one can determine the spin Seebeck coefficient
\beq\label{Ss}
S_s=-\frac{V_s}{\theta}\bigg|_{I_c=0,I_s=0}=\frac{L_{Q}^{\up}}{G_{A}^{\up}+G_{Q}^{\up}}-\frac{L_{Q}^{\down}}{G_{A}^{\down}+G_{Q}^{\down}}\,,
\edq
which measures the generated spin voltage from the application of a temperature difference $\theta$
when both the charge and the spin currents vanish. These conditions give a constraint for the applied voltage. Alternative definitions of $S_s$ are also possible~\cite{mis15}. Here, we have chosen the condition $I_c=0$ and $I_s=0$ because it is a natural extension to the theoretical proposal of charge Seebeck case in an open circuit~\cite{mis14}.

Another important aspect of spin transport is the dependence of the thermopower on the magnetization
or the applied magnetic field~\cite{wal11}. We can then define the magneto-Seebeck coefficients $MS_p$ and $MS_Z$,
which measure variations in the thermopower when $p$ and $\Delta_Z$ are reversed, respectively:
\begin{align}
MS_p&=S(p)-S(-p)\,,\label{MSp}\\
MS_Z&=S(\Delta_Z)-S(-\Delta_Z)\,.\label{MSZ}
\end{align}

\begin{figure}[t]
\centering
  \begin{centering}
    \includegraphics[width=0.45\textwidth]{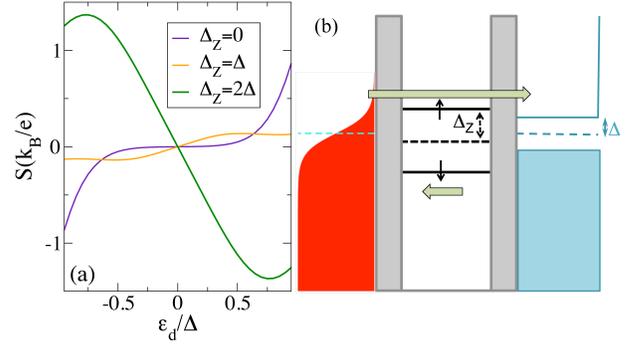}
  \end{centering}
  \caption{(Color online) (a) Thermopower (charge Seebeck coefficient) $S$ versus the level position $\varepsilon_d$ at $p=0$ and background temperature $k_BT=0.2\Delta$ for several values of the Zeeman splitting $\Delta_Z$. We use $\Gamma_F=0.1\Delta$ and $\Gamma_S=0.5\Delta$, i.e., the ferromagnet is the probe terminal, here and throughout the paper.
  (b) Energy diagram for a resonant tunneling double barrier (gray areas) system (the dot) coupled between the normal metal (left) and a superconducting reservoir (right) with $\varepsilon_d<0$ (dashed black line), $E_F=0$ (dashed blue line),
  $\Delta_Z=2\Delta$ (solid black lines) and $p=0$ (normal case). The hot metal displays a thermally smeared
  state distribution (red). At very low temperature, the cold superconductor (blue) has empty quasiparticle states above the gap $\Delta$.
  The filled states below $\Delta$ are represented with solid blue. Note that the current due to temperature excited electrons from the left electrode (right arrow) is much larger than the opposite
  flow of electrons (left arrow), which
  is mostly blocked by filled states in the normal lead for $\varepsilon_d<E_F$. Asymmetric size of the left and right arrows indicates the spin-dependent particle-hole asymmetry generated by gating the dot and reinforced by Zeeman splitting. Accordingly, the net thermoelectric current due to a temperature difference is large.}
  \label{Fig2}
\end{figure}

\section{Results and discussion}\label{results}

The above formulas are general and can be applied to a variety of situations.
In this section, we consider the case where the dot is more strongly coupled to the superconducting
lead ($\Gamma_F=0.1\Delta$ and $\Gamma_S=0.5\Delta$). This amplifies the effects discussed
below but qualitatively the physics remains the same if the opposite situation (ferromagnetic
dominant case) is considered.

Our results rely on the breaking of particle-hole symmetry. This can be done in three different ways. First, the gating of the quantum dot energy level $\varepsilon_d$ away from the Fermi energy breaks the symmetry and thus creates a finite thermoelectric signal. Second, the Zeeman field splitting $\Delta_Z$ can enhance the effect of gate potential if one of the spin split levels is shifted out of the gap region. Third, the spin polarization $p$ of ferromagnetic lead causes an asymmetry in the electron and hole transport in case of spin-split dot levels.

Figure~\ref{Fig2}(a) presents the charge thermopower as a function
of the dot level position $\varepsilon_d$ for various Zeeman splittings $\Delta_Z$ and magnetization $p=0$. At the symmetric point when the dot
level is aligned with the Fermi energy, the Seebeck coefficient $S$ vanishes independently
of $\Delta_Z$. We observe that $S$ is close to zero for most values of the dot level within the gap
if both $p$ and $\Delta_Z$ are zero. As discussed above, we thus need to break the particle-hole symmetry
in the system by the simultaneous application of nonzero $\Delta_Z$ and $\varepsilon_d$.
When $\Delta_Z$ is of the order of the superconducting gap, the thermopower increases
as compared with the $\Delta_Z=0$ case but the effect is more dramatic when
$\Delta_Z=2\Delta$. The thermopower attains large negative (positive) values for
positive (negative) $\varepsilon_d$. This can be understood from the energy
diagram shown in Fig.~\ref{Fig2}(b) where we depict the case of a large Zeeman splitting
and a negative $\varepsilon_d$. Due to the energy dependence of the superconducting
density of states, the level $\varepsilon_{d\downarrow}$ lies in a region with a small
number of available states. Transport then takes place mainly across the upper
$\varepsilon_{d\uparrow}$ level. Since this leads to a positive thermocurrent
[large arrow in Fig.~\ref{Fig2}(b)],
the definition of Eq.~\eqref{S} implies that the thermopower is thus positive.
Remarkably, our device shows great values of $|S|$ even if $p=0$, which clearly differs
from ferromagnet-superconductor junctions where the thermoelectric effect
is predicted to vanish if $p=0$~\cite{oza14}.

\begin{figure}[t]
\centering
  \begin{centering}
    \includegraphics[width=0.45\textwidth]{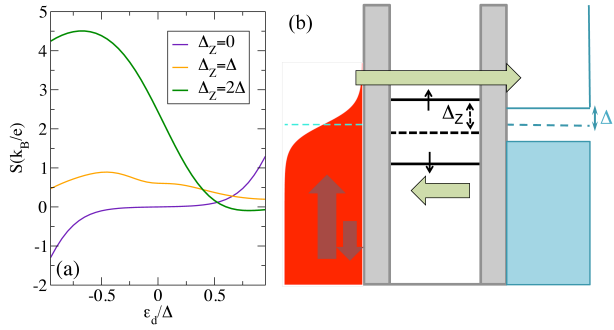}
  \end{centering}
  \caption{(Color online) (a) $S$ versus $\varepsilon_d$ at $p=0.9$ and $k_BT=0.2\Delta$ for several $\Delta_Z$.
  (b) Energy diagram analogous to Fig.~\ref{Fig2}(b) but with a ferromagnetic electrode (different spin populations
  are indicated with vertical arrows). Since more electrons with spin up are now available for tunneling, the thermocurrent
  increases as compared to the normal case in Fig.~\ref{Fig2}(b).}
  \label{Fig3}
\end{figure}

The case of nonzero magnetization is illustrated in Fig.~\ref{Fig3}(a).
The thermopower is largely enhanced when the Zeeman splitting increases.
In comparison with Fig.~\ref{Fig2}(a) we find that the combination
of magnetic fields and ferromagnetic contacts leads to values of $S$
of the order of $4$ (in units of $k_B/e=86$~$\mu$V/K)
for certain values of the dot level. This increase is clarified in Fig.~\ref{Fig3}(b).
Because there exist more electrons with spin $\uparrow$ in the left lead,
the thermocurrent increases since more electrons are able to tunnel through
the upper dot level and the thermopower, which is proportional to the thermoelectric
conductance, thus grows [cf. the case $p=0$ in Fig.~\ref{Fig2}(b)].
On the other hand, the relatively small
values of $S$ for positive $\varepsilon_d$ can be explained from a compensation
effect. If $\varepsilon_d>0$ the energetically favorable channel is $\varepsilon_{d\downarrow}$
but few electrons with spin $\downarrow$ are available as $p=0.9$.
Therefore, $I_c$ decreases and the generated thermovoltage is low.
This demonstrates that thermoelectric effects can be highly tunable
by changing the gate potential applied to the dot.

\begin{figure}[t]
\centering
  \begin{centering}
    \includegraphics[width=0.35\textwidth, angle=-90]{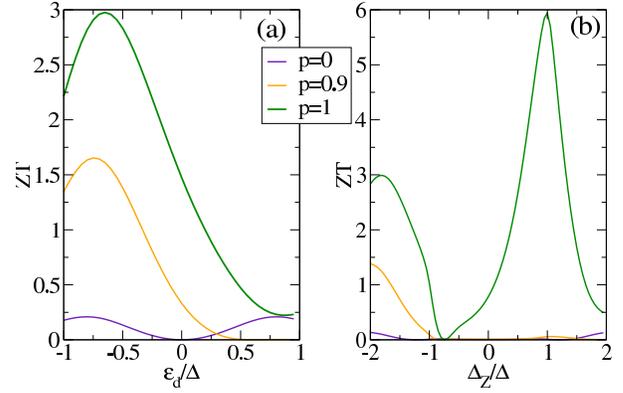}
  \end{centering}
  \caption{(Color online) Thermoelectric figure of merit $ZT$ versus (a) $\varepsilon_d$ at $\Delta_Z=2\Delta$ and (b) $\Delta_Z$ at $\varepsilon_d=0.5\Delta$ for several polarization values of the ferromagnetic electrode at $k_BT=0.2\Delta$.}
  \label{Fig4}
\end{figure}

The thermoelectric figure of merit calculated from Eq.~\eqref{ZT}
is displayed in Fig.~\ref{Fig4}(a). Its behavior follows the thermopower
properties discussed above. $ZT$ increases for negative $\varepsilon_d$
when both $p$ and $\Delta_Z$ are positive and large.
The exact value of $ZT$ can be also tuned at a fixed position
of the dot level. This is illustrated in Fig.~\ref{Fig4}(b),
where the figure of merit reaches very high values
as a function of the applied magnetic field, especially
in the half-metallic case ($p=1$).
Our results thus show that a F-D-S device may act as an efficient
waste heat-to-electric energy generator.
Furthermore, this system could also be useful for cooling applications
at very low temperature since Peltier and Seebeck effects are reversible.

\begin{figure}[t]
\centering
  \begin{centering}
    \includegraphics[width=0.35\textwidth,angle=-90]{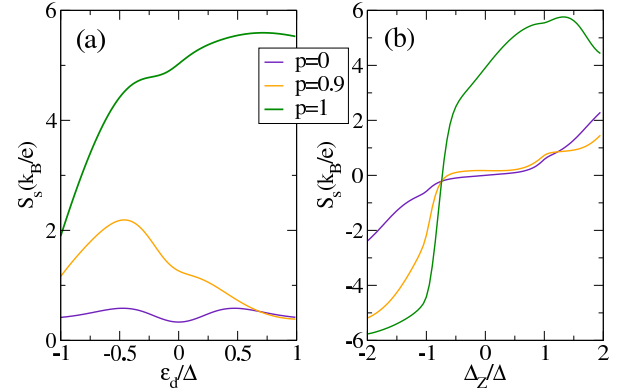}
  \end{centering}
  \caption{(Color online) Spin Seebeck coefficient $S_s$ versus (a) $\varepsilon_d$ at $\Delta_Z=\Delta$ and (b) $\Delta_Z$ at $\varepsilon_d=0.5\Delta$ for several polarization values of the ferromagnetic electrode at $k_BT=0.2\Delta$.}
  \label{Fig5}
\end{figure}

The spin Seebeck coefficient $S_s$ is calculated from Eq.~\eqref{Ss}. In Fig.~\ref{Fig5}(a)
we show the results as a function of the dot level for fixed Zeeman splitting
and increasing values of the ferromagnet polarization. We find that $S_s$ increases
with $p$ and reaches high values in the half-metallic case.
The spin Seebeck coefficient is always positive because in the right-hand side of Eq.~\eqref{Ss}
the first term, which corresponds to electrons with spin $\uparrow$, dominates over
the second term. In fact, $L_Q^\uparrow$ grows as $p$ increases since more electrons
with spin $\uparrow$ are available for tunneling whereas at the same time $L_Q^\downarrow$ 
decreases. In the analysis of $S_s$ as a function of the Zeeman splitting keeping $\varepsilon_d$
constant [Fig.~\ref{Fig5}(b)], we obtain an interesting change of sign. For positive $\Delta_Z$ the spin Seebeck
coefficient is positive for the reasons discussed above. However, for sufficiently negative
values of $\Delta_Z$, $S_s$ becomes negative since now $\varepsilon_{d\uparrow}$ lies
below the Fermi energy and $L_Q^\uparrow$ then changes sign while $L_Q^\downarrow$
is still close to zero for large values of $p$. In any case, the results for $|S_s|$ are large
provided the Zeeman splitting is of the same order as the superconducting gap.

\begin{figure}[t]
\centering
  \begin{centering}
    \includegraphics[width=0.35\textwidth,angle=-90]{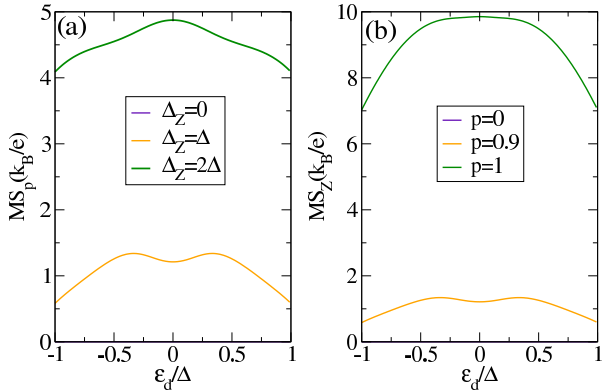}
  \end{centering}
  \caption{(Color online) Magneto-Seebeck coefficients (a) $MS_p$ at $p=0.9$ and (b) $MS_Z$ at $\Delta_Z=\Delta$ as a function of the dot level position for $k_BT=0.2\Delta$.}
  \label{Fig6}
\end{figure}

Both the magneto-Seebeck coefficients $MS_p$ and $MS_Z$ as defined in Eqs.~\eqref{MSp} and~\eqref{MSZ}
are respectively plotted in Figs.~\ref{Fig6}(a) and~\ref{Fig6}(b) as a function of the dot level position.
For $MS_p$ we fix $p=0.9$ and vary the Zeeman splitting while for $MS_Z$
we set $\Delta_Z=\Delta$ and change the ferromagnetic polarization.
Strikingly, all curves show a characteristic symmetry with regard to the Fermi energy.
In the case of $MS_p$ this is understood from the relation $S(p,\varepsilon_d)=-S(-p,-\varepsilon_d)$
for a given $\Delta_Z$. Physically, it means that an electron-hole transformation that shifts the
dot level with respect to $E_F=0$ and simultaneously reverses the ferromagnetic polarization
induces a thermocurrent with the opposite sign.  This can be seen in Fig.~\ref{Fig2}(a) for $p=0$.
In the case of $MS_Z$, a similar symmetry relation holds, namely,
$S(\Delta_Z,\varepsilon_d)=-S(-\Delta_Z,-\varepsilon_d)$.
Overall, both magneto-Seebeck coefficients increase for larger Zeeman splittings or ferromagnetic
polarizations, in agreement with our previous results.

\section{Conclusions}\label{conclusion}
We have investigated the thermoelectric properties of
a ferromagnetic-quantum dot-superconducting device
in the presence of an external magnetic field applied to the
dot. We have shown that the device develops high values
of the thermopower from the combined effect of spin polarized
tunneling, Zeeman splitting and tuning of the dot level.
Importantly, the thermoelectric conversion is efficient
since the dimensionless figure of merit reaches values
as high as $6$. Moreover, the spin Seebeck effect exhibits
relevant changes as a function of the gate potential
and the magneto-Seebeck coefficient becomes sensitive
with reversals of the magnetization direction or the applied
magnetic field.

Our predictions can be tested with today's
experimental techniques. The quantum dot can be formed
inside a carbon nanotube attached to ferromagnetic and
superconducting contacts. Another possibility is to use
a nanowire deposited onto the ferromagnet and the superconductor~\cite{hof10}.
This system is especially appealing since, e.g., InSb nanowires have large effective
$g$ factors. Then, for $\Delta_Z=g\mu_BB/2=2\Delta$ we estimate a magnetic field $B\simeq 5$~T
using $g=40$ and $\Delta=3$~meV for Nb~\cite{den12}. We then need a superconducing lead
with high critical field $B_c$. Nb compounds precisely show the property that $B_c>B$.
Then, we can neglect the dependence of $\Delta$ on the magnetic field to a first approximation.
One could also envision a self-assembled quantum dot connected
to two electrodes as in Ref.~\cite{kat10} or a break junction comprising
a C$_{60}$ molecule~\cite{pas04,win09}.
The ferromagnetic electrode can be Joule heated with a slowly time-dependent
electric current with zero average and thus leading to a temperature
shift across the junction~\cite{sta93,fah13}. Most of the applied temperature
bias drops at ferromagnetic intereface~\cite{kol15} and thus the superconducting
temperature (and thereby its energy gap) is largely unaffected by the thermal gradient.
The detection of the spin bias can be done using the inverse spin Hall effect~\cite{sai06}.
Finally, the magneto-Seebeck effect needs the magnetization to be switched,
which can be accomplished, e.g., by employing an external magnetic field that can be really
small for soft ferromagnets.

Our work raises two important questions. First, what is the role of electron-electron
interactions? Our theory assumes a single-level dot large enough that Coulomb repulsion
is negligible. The results would also hold for strongly coupled quantum dots since
in this case the charging energy $U$ is smaller than the tunneling broadening and electronic interactions
can be safely disregarded. However, small dots usually have large $U$ and Coulomb
blockade effects become dominant. It would be even possible to explore strongly correlated
phenomena such as the Kondo effect~\cite{kondo1,kondo2}. An Anderson-like Hamiltonian should be then
used. The second question concerns the role of higher-order terms in the current--voltage
or current--temperature characteristics~\cite{san13,hwa13} and heat~\cite{yam15}.
Even if larger temperature biases do not contribute
to the Andreev current there appear cross terms that mix voltage and temperature differences~\cite{hwa15}.
In that case, one should resort to differential thermopowers and the results might change
significantly. Our work is thus a first approach to a problem with fertile ramifications.

\begin{acknowledgments}
This research was supported by MINECO under Grant No.\ FIS2014-52564 and the Korean NRF-2014R1A6A3A03059105.
\end{acknowledgments}

\appendix

\end{document}